\documentclass[fleqn,twoside]{article}
\usepackage{espcrc2}

\usepackage{graphicx}
\usepackage[figuresright]{rotating}

\newcommand{\bi}{\begin{itemize}}
\newcommand{\ei}{\end{itemize}}

\newcommand{\dzero}	{D\O}
\newcommand{\AmS}{{\protect\the\textfont2
  A\kern-.1667em\lower.5ex\hbox{M}\kern-.125emS}}

\title{Optimized Neural Networks to Search for Higgs Boson Production at 
the Tevatron.}

\author{E.~Boos, L.~Dudko %
\address{Scobeltsyn Institute of Nuclear Physics of
Moscow State University,\\ Vorobyevy Gory, 119899, Moscow, Russia}%
}
       
\begin{document}
\begin{abstract}
An optimal  choice of proper kinematical variables is one of the main 
steps in using neural networks (NN) in high energy physics.  
Our method of the variable selection is based on the analysis 
of a structure of Feynman diagrams  
(singularities and spin correlations) contributing to  
the signal and background processes. An application of this method 
to the Higgs boson search at the Tevatron leads to an improvement 
in the NN efficiency by a factor of 1.5-2 in comparison to previous 
NN studies.
\vspace{1pc}
\end{abstract}
\maketitle
\section{The basic idea}
In High Energy physics a discrimination between a signal and its 
corresponding backgrounds by Neural Networks (NN) is
especially remarkable when the data statistics are limited.
In this case it is important to optimize all steps of the analysis.
One of the main questions which arises in the use of NNs 
is which, and how many variables should be chosen
for network training in order to extract 
a signal from the backgrounds in an optimal way.
The general problem is rather complicated and finding a solution
depends on having a concrete process for making the choice, because
 usually it takes a lot
of time to compare results from different sets of variables.     

One observation which helps in making the best choice of the most sensitive
variables  is to study the singularities in Feynman diagrams of the 
processes.
Let us call those kinematic variables 
in which singularities occur as "singular variables".
What is important
to stress here is that most of the rates for both the signal and for
the backgrounds come from the integration over the phase space region 
close to these singularities. 
One can compare the lists of singular variables and the positions of the 
corresponding singularities in Feynman diagrams for the 
signal process and for the  backgrounds. 
It is obvious that if some 
of the singular variables are different or the positions of 
the singularities
are different for the same variable 
for the signal and for the backgrounds the
corresponding  distributions will differ most strongly. 
Therefore, if one uses all such 
singular variables in the analysis, then the largest part of the phase space
where the signal and backgrounds differ most will be taken
into account.
One might think that it is not a simple task 
to list all the singular variables when the phase space
is very complex, for instance, for 
reactions with many particles involved.
However, in general, all singular variables can be of
only two types, either s-channel: $M_{f1,f2}^2 = (p_{f1} + p_{f2})^2$, 
where $p_{f1}$ and $p_{f2}$ are the four
momenta of the final particles  $f1$ and $f2$
or t-channel: $\hat{t}_{i,f} = (p_f-p_i)^2$, where $p_f$  
and  $p_i$ are the momenta of the final particle (or cluster)
and the initial parton. For the $\hat{t}_{i,f}$ all the needed variables
can be easily found in 
massless case: $\hat{t}_{i,f} = - \sqrt{\hat{s}} e^{Y} p_T^f e^{-|y_f|}$, 
where $\hat{s}$ 
is the total invariant mass of the
produced system, and {\it Y} is the rapidity of the total system (rapidity
of the center mass of the colliding partons), $p_T^f$ and $y_f$ 
are transverse momenta and pseudorapidity of the
final particle {\it f}.
The idea of using singular variables as the most discriminative ones is  
described in~\cite{BDO} and the corresponding method was demonstrated in 
practice in~\cite{aihenp}.

   Singular variables correspond to the structure of the denominators
 of Feynman diagrams. Another type of interesting variables corresponds to
the numerators of Feynman diagrams and reflects the spin effects and 
the corresponding
 difference in angular distributions of the final particles. 
In order to discriminate between a signal and the backgrounds, 
one should choose in addition to singular variables mentioned above 
those angular variables whose distributions are different 
for the signal and backgrounds. The set of these singular and 
angular variables will be the most efficient set for a NN analysis.

The third type of useful variables which we called "Threshold"
variables are related to the fact that various signal and background
processes may have very different thresholds. Therefore the distributions
over such kind of variables also could be very different keeping in mind
that effective parton luminosities depend strongly on 
$\hat{s}$. The variable $\hat{s}$ would be a very efficient variable of 
that kind. However, the problem is that in case of neutrinos in the final 
state one can not measure $\hat{s}$ and should use the effective
$\hat{s}$ which is reconstructed by solving t-,W-mass equations 
for the neutrino longitudinal momenta. That is why we propose to use
not only the effective variable $\hat{s}$ but the variable
$H_T^{jets}$ as well.  

To apply the method it is important to use a proper Monte-Carlo model 
of signal and background events which includes all needed spin 
correlations between production and decays. 
For the following analysis we have calculated the complete tree level
matrix elements for the background processes with all decays and 
correlations by means of the CompHEP program~\cite{comphep}.
The corresponding events are available at the FNAL Monte-Carlo events 
database~\cite{mcdb}.

\section{Applying the method}
The present estimation of the expected sensitivities for the light Higgs 
boson 
search at the Tevatron by means of NNs is given 
in~\cite{prd_pushpa}. Based on the method 
described above we improve the efficiency of the NN technique. 
In the analysis we choose the Higgs boson mass to be
$M_H=115$~GeV. We model the detector smearing by the SHW 
package~\cite{SHW}.

\begin{figure}[htb]
  \hspace*{-1cm}
\begin{minipage}{.49\linewidth}
\includegraphics[width=12pc,height=11pc]{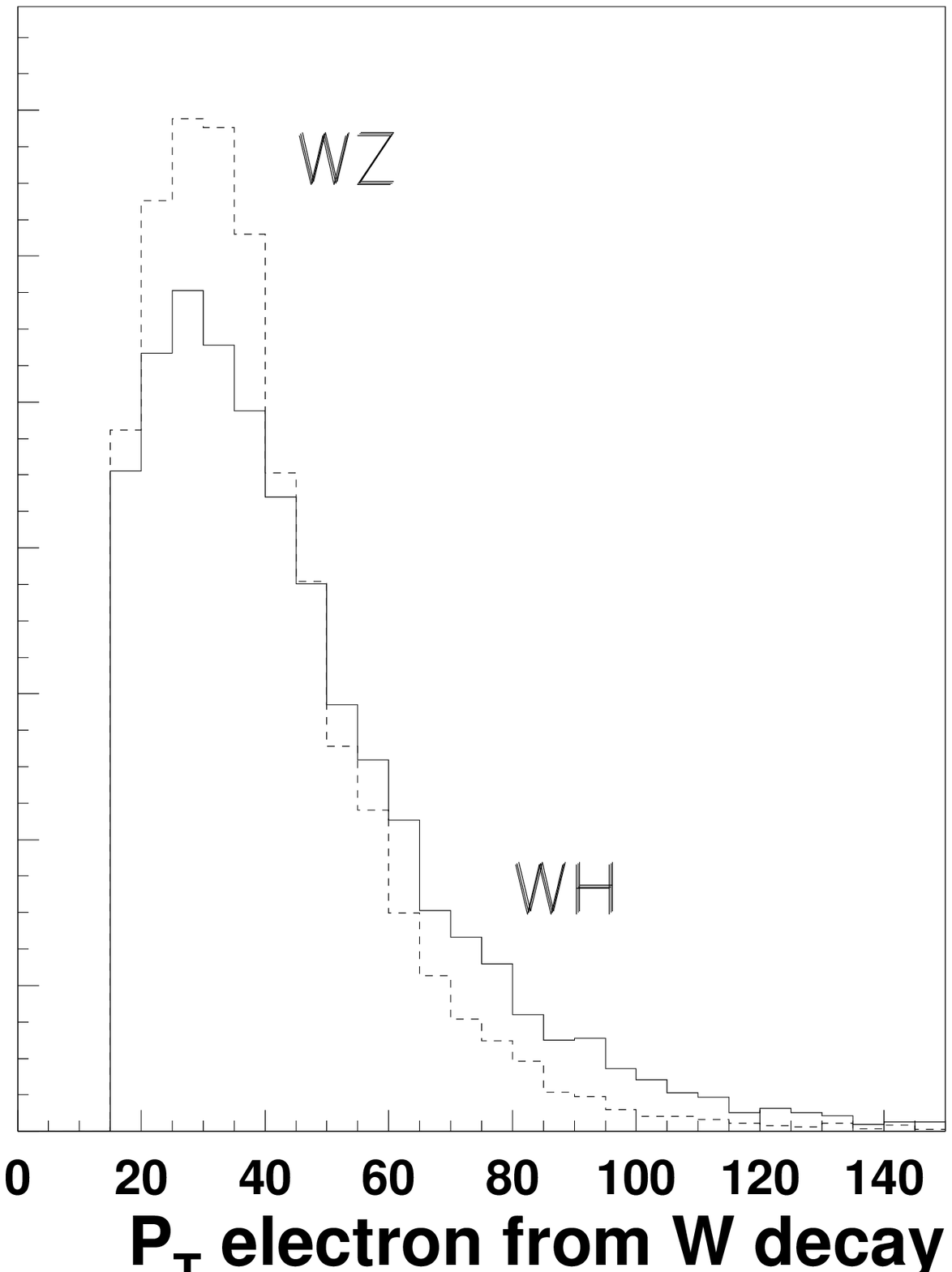}
\end{minipage}
\begin{minipage}{.49\linewidth}
\includegraphics[width=12pc,height=11pc]{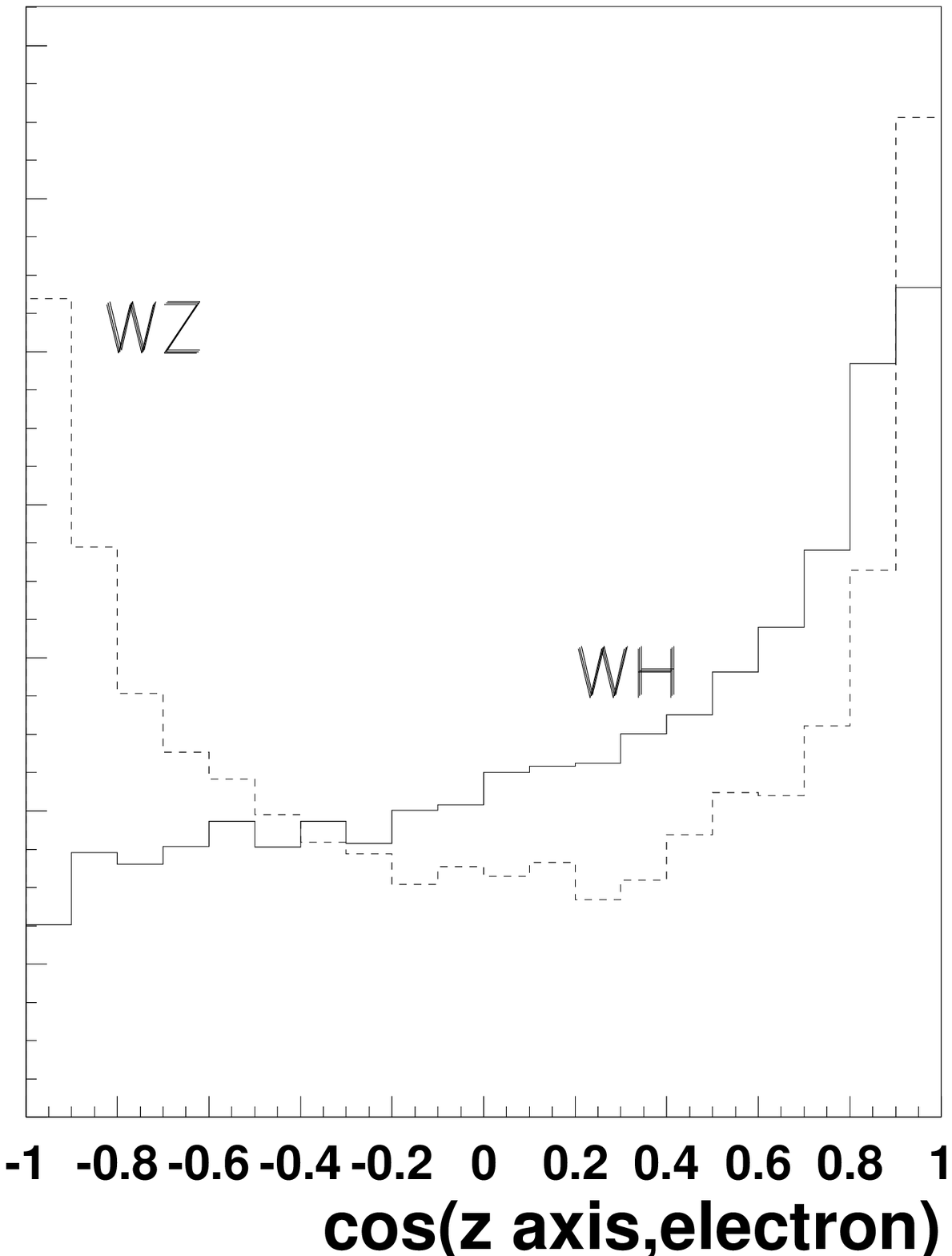}
\end{minipage}
\caption{Examples of the old kinematic variable (left plot) and the new one (right
plot)}
\label{fig:pt}
\end{figure}

First of all we exclude ineffective variables
from the old set~\cite{prd_pushpa}, like $P_T^e$ from the $W$-boson (shown 
at 
the left plot in Fig.~\ref{fig:pt}). After the corresponding analysis of 
Feynman diagrams and comparison of kinematical distributions we added the 
new variables for NN training. The example distribution for the new 
variable 
($cos(z_{axis},e)$) is shown in the right plot of Fig.~\ref{fig:pt}.
  At the next step we constructed the set of NNs for pairs 
of the signal (WH) and each 
of the background from the complete set of principle backgrounds 
($Wb\bar b,\ WZ,\ t\bar t,\ tb(j)$). 

The standard steps of NN
training were used for the NNs with the old set of input variables and 
with the new one.
Efficiencies of networks with
different sets have been compared 
based on the criteria that for the better net the  ``Error function'' 
$ E = \frac{1}{N}\sum_{i=1}^{N}(d_i-o_i)^2$, where $d_i$ and $o_i$ 
are the desired and real
outputs of the net and $N$ is the number of test events, is smaller.
Two examples of distributions you can see in Fig~\ref{fig:e} 
for the $WH-t\bar t$
network (left plot) and $WH-WZ$ network (right plot). One can see  
a significant improvement for the networks with new input sets in 
comparison with old sets
 of variables, since the corresponding curves of the error 
function are significantly
 lower. 
 \section{Results}
 Based on the described method we have constructed the new NNs to search 
for a light
 Higgs boson at the Tevatron. After checking the improvement in efficiency
 of new networks we recommend the new sets of input variables for NNs, which are
 shown below:\\
\begin{itemize}
\item 
$Wb\bar b$ -- $WH$ \\
NN: 
 $M_{b\bar b},\ P_{T}^{b1},\ P_{T}^{b2},\ P_{T}^{bb},\
\hat{s},\ H_T^{jets},$\\ \hspace{0.3cm} $cos(b1,b2)|_{lab},\ cos(b1,b1b2)|_{b1b2}$\\ 
\item 
$WZ$ -- $WH$ \\
NN: 
 $M_{b\bar b},\ P_{T}^{b1},\ P_{T}^{b2},\  H_T^{jets},$\\
$cos(b1,b2)|_{lab}, \ Q\times cos(z,b1)|_{lab},\ cos(W,e)|_{W}$\\  
\item  
$t\bar t$ -- $WH$ \\
NN: 
 $M_{b\bar b},\ M_{Wb},\ \hat{s},\ M_{Wjets-b},\
H_T^{jets},$\\ $ Q\times cos(\psi_{axis},e)|_{top},\  cos(b1,b1b2)|_{b1b2}$\\
\item  
$tbj,\ tb$ -- $WH$ \\
NN: 
 $M_{b\bar b},\ M_{Wb},\ P_{T}^{b2},\ \hat{s}, $\\ $M_{Wjets-b},\
P_{T}^{top},\  H_T^{jets},\ cos(z,e)|_{lab},$\\ $ Q\times cos(z,b1)|_{top},\
cos(e,j)|_{top}$\\
\end{itemize}
where there are three types of variables:
\bi 
\item ``Singular'' variables (denominator of Feynman diagrams):\\
   $M_{12}$ is the invariant mass of two particles and/or jets (1 and 2) and 
corresponds to s-channel singularities;\\
   $P_{T}^{f}$ (the transverse momenta of {\it f});\\ 
   $M_{Wjets-b}$ is the invariant mass of the $W$ and all jets except 
the $b$-jet for which the
   $M_t=(p_W+p_b)$ is closest to the top quark mass;\\
\item ``Angular'' variables (numerator of Feynman diagrams, spin effects):
   $cos(b1,b1b2)|_{b1b2}$ means the cosine of the angle between highest $P_T$ 
b-quark 
   and vector sum of the two highest $P_T$ b-quarks in the rest frame of 
these two b-quarks. 
Scalar (Higgs) and vector (gluon, Z-boson) particle decays lead to 
significantly different distributions
on this variable, this is also very much different for the case when b-quarks
come from the decay of top and anti-top quarks; \\
   $cos(b1,b2)|_{lab}$ characterizes how much two b-quarks are 
collinear;\\
   $cos(z,b1)|_{lab}$ and $cos(W,e)|_{W}$ reflect the difference 
in t-channel Z-boson  
   and s-channel Higgs-boson production topologies 
where $_{lab}$ means the laboratory rest frame
   and $z$ means the  z-axis;\\
   $cos(\psi_{axis},e)|_{top}$~\cite{tt_spin} and 
$cos(e,j)|_{top}$~\cite{single_top_spin} 
   are the top quark spin correlation 
   variables used in the analysis of the top quark 
pair and single production, the 
   lepton charge  
 $Q$ is added here to take uniformly into account the 
electron and the positron contributions 
 from the $W$-boson decays.
\item ``Threshold'' variables. 
As explained above  
 the $\hat{s}$ and $H_T^{jets}$ variables are used in our analysis. 

\ei

\begin{figure}[htb]
  \hspace*{-0.9cm}
\begin{minipage}{.49\linewidth}
\includegraphics[width=12pc,height=11pc]{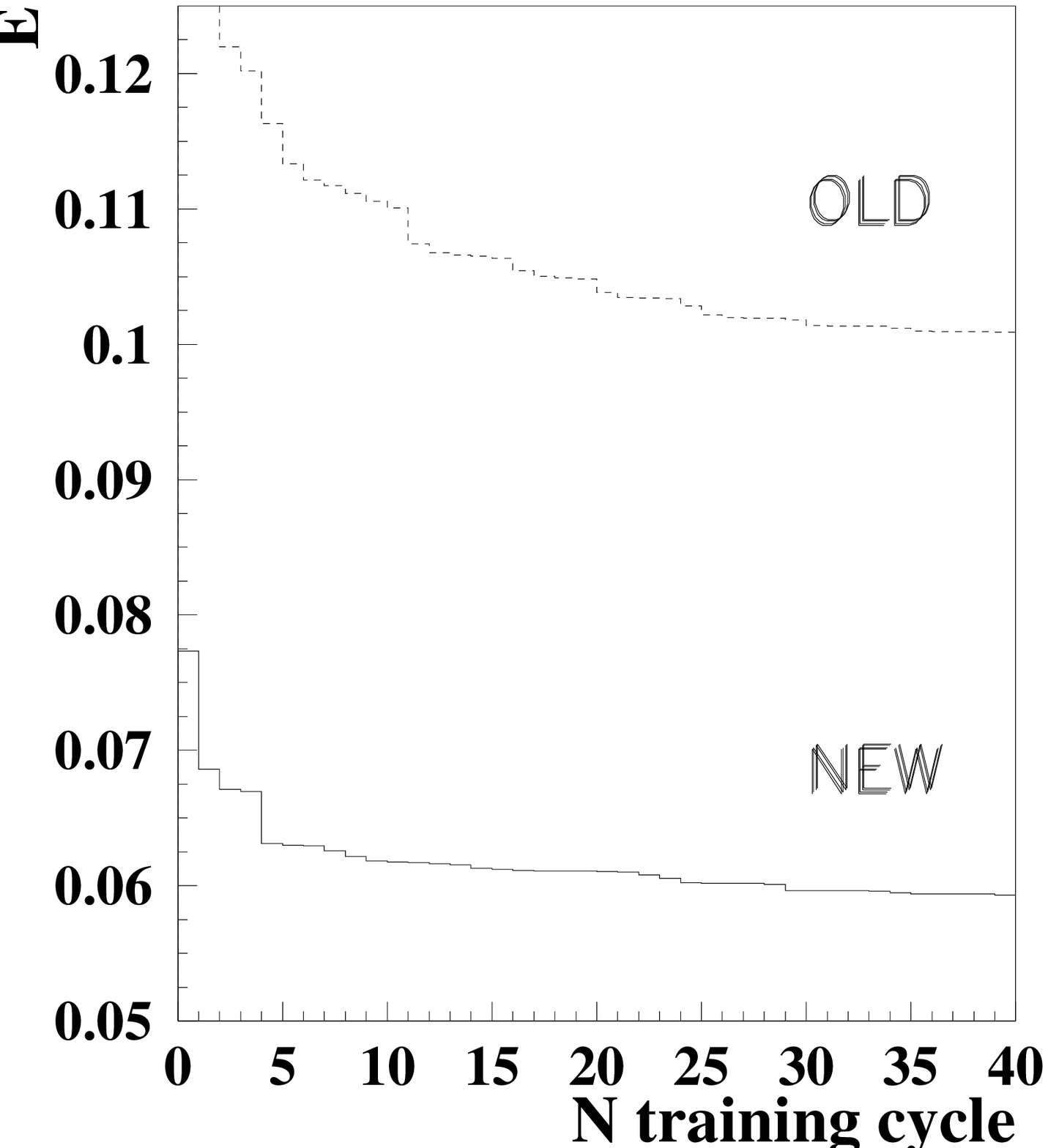}
\end{minipage}
\begin{minipage}{.49\linewidth}
  \hspace*{0.5cm}
\includegraphics[width=12pc,height=11pc]{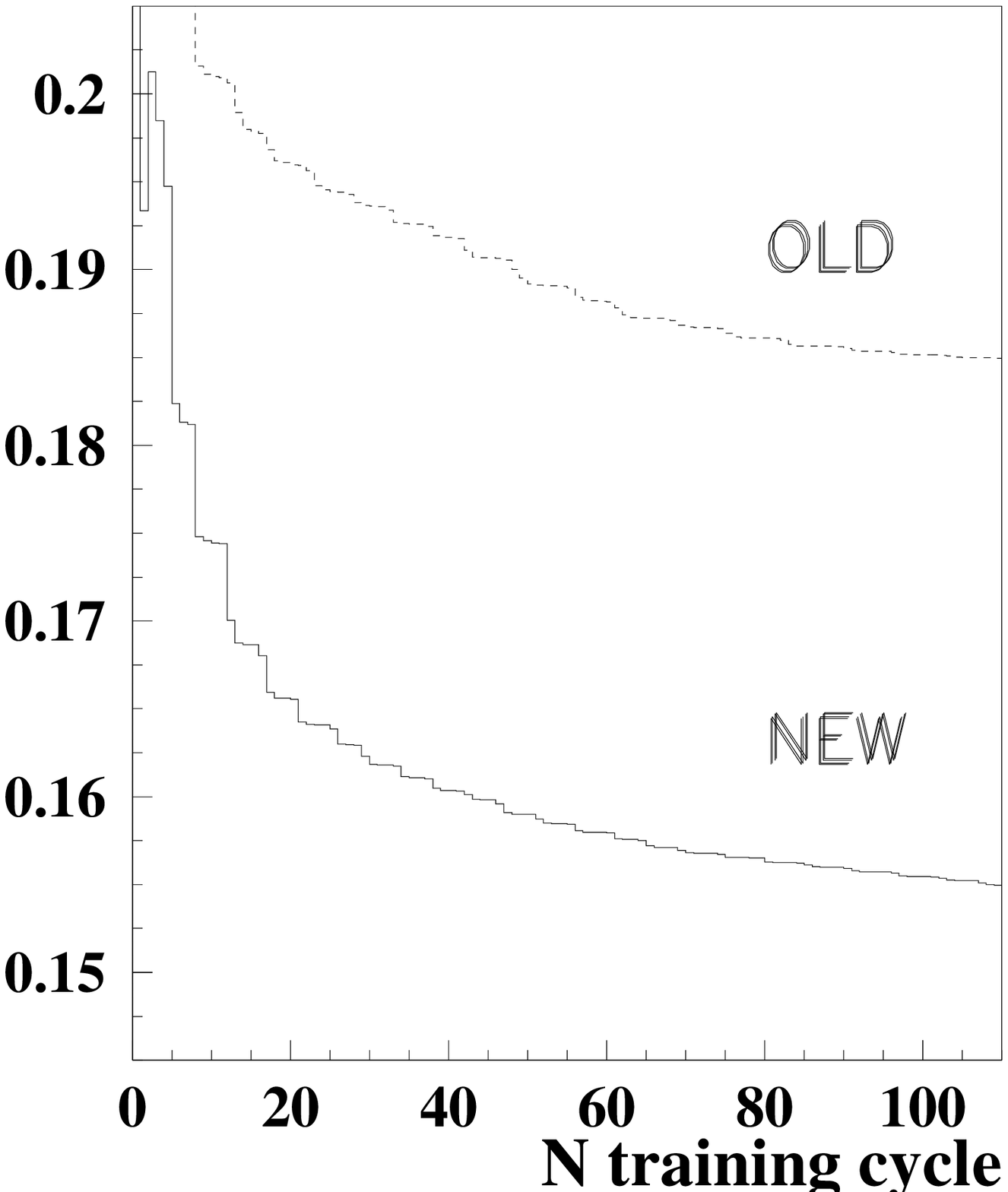}
\end{minipage}
\caption{NN Error function for the $WH-t\bar t$
(left plot) and $WH-WZ$ networks (right plot).}
\label{fig:e}
\end{figure}

As one can see from the Fig.\ref{fig:e} using the new NN variables
allows to improve the NN efficiency by a factor of 1.5-2 depending on the 
background process.
It will lead to corresponding improvement in prospects to find
a light Higgs at the Tevatron. However, one needs to take into account
the ZH production channel as well as a number of detector efficiencies
in order to predict a realistic discovery limit.
\section*{Acknowledgements}
The work was partly supported by
the INTAS 00-0679, CERN-INTAS 99-377, YSF 02/239 and 
Universities of Russia UR.02.03.002 grants.
E.~Boos thanks the Humboldt Foundation for the Bessel Research Award.

\end{document}